\documentclass[conference]{IEEEtran}
\IEEEoverridecommandlockouts
\usepackage{algorithm}
\usepackage{hyperref}
\usepackage{xcolor}
\usepackage{float}
\usepackage{optidef}
\usepackage{cases}
\usepackage{amsmath}
\usepackage{amsfonts}
\usepackage{mathtools, nccmath}
\usepackage{ulem}
\def\BibTeX{{\rm B\kern-.05em{\sc i\kern-.025em b}\kern-.08em
    T\kern-.1667em\lower.7ex\hbox{E}\kern-.125emX}}
\DeclarePairedDelimiter{\nint}\lfloor\rfloor
\usepackage{titlesec}
\titlespacing*{\section}{0.1pt}{0.1\baselineskip}{0.1\baselineskip}
\usepackage[nodisplayskipstretch]{setspace}
\begin{document}

\title{INR-MDSQC: Implicit Neural Representation Multiple Description Scalar Quantization for robust image Coding\\
\thanks{This project is funded by the Région Sud, France}
}

\author{\IEEEauthorblockN{Trung Hieu Le, Xavier Pic, Marc Antonini, member IEEE}\\
\textit{I3S laboratory, Côte d’Azur University CNRS,UMR 7271, Sophia Antipolis, France}}
\maketitle

\begin{abstract}
Multiple Description Coding (MDC) is an error-resilient source coding method designed for transmission over noisy channels. We present a novel MDC scheme employing a neural network based on implicit neural representation. This involves overfitting the neural representation for images. Each description is transmitted along with model parameters and its respective latent spaces. Our method has advantages over traditional MDC that utilizes auto-encoders, such as eliminating the need for model training and offering high flexibility in redundancy adjustment. Experiments demonstrate that our solution is competitive with autoencoder-based MDC and classic MDC based on HEVC, delivering superior visual quality.
\end{abstract}


\section{Introduction}
Multiple Description Coding (MDC) has been studied for many years. In \cite{DBLP:journals/spm/Goyal01a}, the authors presented an efficient source coding solution able to manage packet errors, random bit errors and routing delays. MDC for image encoding involves encoding multiple representations of an image; if one is lost or corrupted during transmission, the remaining descriptions can still be used to reconstruct the original image with some quality degradation.

Classic MDC methods have typically dealt with some problems. The first application with a scalar quantizer was proposed in \cite{DBLP:journals/tit/Vaishampayan93}, where the index assignment refers to the process of mapping from the source to a set of output descriptions to achieve the best rate, redundancy, and distortion trade-off. This problem is complex. The wavelet transform MDC is based on \cite{DBLP:conf/icip/PereiraAB03}, where authors confront issues of quantization and redundancy index assignment, and attempt to solve the problem of non-linearity during optimization by modeling each subband a Gaussian model. However, this model has limited accuracy at lower rates, and its complexity is very high. Standard-compliant MDC methods such as HEVC \cite{le2023multiple,DBLP:journals/tcsv/TilloGO08,DBLP:journals/jvcir/WangCZC22} can achieve high performance with low latency. However, rate distortion control is carried out with an empirical formula that is based on linear regression, which limits the quantization range and thereby constrains their performance.

Recent research has indicated the potential use of neural networks for image compression \cite{DBLP:conf/pcs/BalleLS16,DBLP:conf/iclr/TheisSCH17,DBLP:conf/cvpr/MentzerATTG18}, but a few work have applied it to MDC. The most recent applications for MDC are \cite{DBLP:journals/tcsv/ZhaoBWZ19,DBLP:journals/mta/ZhaoZBWZ22}, which employ Generative Networks and Compressive Autoencoders. However, these methods' drawbacks include the requirement of a high computing capacity for the training process. Furthermore, the training process must be performed with very large datasets to be efficient. This is even more challenging in the MDC context due to the redundancy adaptation mechanism, which requires retraining the model.

In recent image compression research using neural networks, the so-called {\it Implicit Neural Representation} (INR), the neural network learns to represent an image implicitly through its weights, a coordinate map, and possibly a latent space \cite{sitzmann2019siren,DBLP:conf/eccv/StrumplerPYGT22}. More recently, the Coordinate-based Low Complexity Hierarchical Image Codec (COOL-CHIC) framework \cite{arxiv:ladune2023coolchic} has achieved performance close to the state of the art of the compressive autoencoder presented in \cite{DBLP:conf/pcs/BalleLS16}, without the need for a training process.
In this paper, we propose the Implicit Neural Representation Multiple Description Scalar Quantization Codec (INR-MDSQC) method based on the COOL-CHIC architecture. The advantages of the proposed solution are:  no need for model training, high performance and flexible redundancy tuning.

In the rest of the paper, we first formulate in section \ref{sec:MDCpbstatement} our MDC problem using this network. Then, we present in section \ref{sec:synthesis} and \ref{sec:autoregressive} the detailed architectures of the synthesis model and the auto-regressive model, both of which are optimized during training. Following that, we discuss in section \ref{sec:modelquantization} the post-training quantization process designed for precision reduction. Lastly, in section \ref{sec:bsstructure} we outline the bitstream's organization and the decoding process. Finally, we show the experimental result in section \ref{sec:ExpResult} and conclude the paper in section \ref{sec:conclusion}.

\section{Proposed Method}
\subsection{Multiple description problem statement \label{sec:MDCpbstatement}}

Inspired by the COOL-CHIC framework \cite{arxiv:ladune2023coolchic}, we propose an overfitted INR-MDSQC network based on hierarchical latent scalar quantization. The architecture of the INR-MDSQC has three main components as illustrated in figure \ref{fig:globalarch}:
\begin{figure}[htp]
    \centering
    \includegraphics[width=\linewidth]{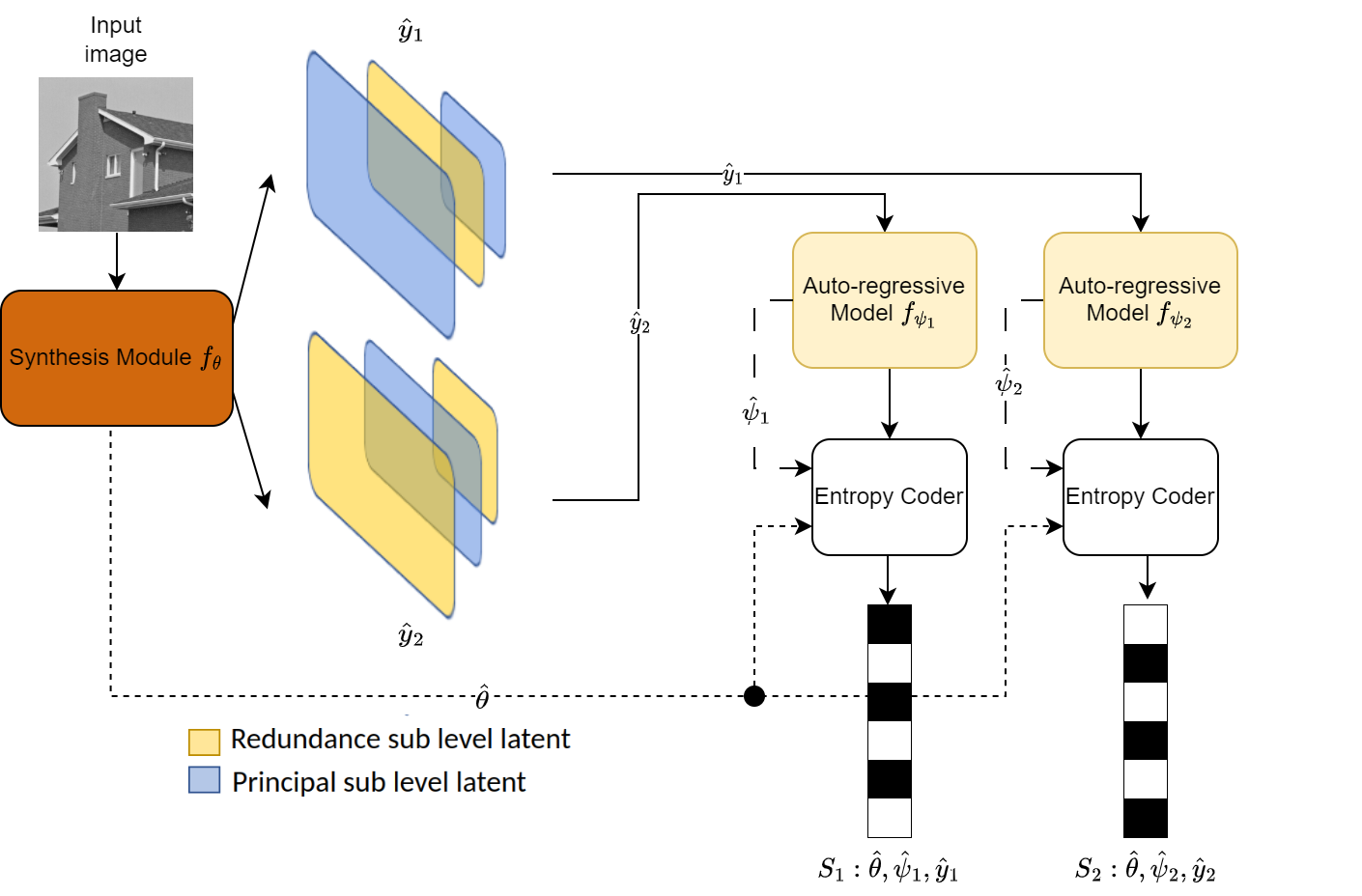}
    \caption{\footnotesize \textbf{INR-MDSQC}: Given an image, the synthesis model ($f_\theta$) divides it into two latent spaces, $\mathbf{\hat{y}_1}$ and $\mathbf{\hat{y}_2}$. Each latent space is then compressed based on the probability estimates derived from the auto-regressive model. The entropy coder proceeds to compress both the model parameters ($\theta$, $\psi$) and the pixels in the latent space to generate two descriptions $S_1$ and $S_2$}
    \label{fig:globalarch}
\end{figure}
\begin{enumerate}
    \item \textbf{Two sets of discrete hierarchical latent spaces}: $\mathbf{y_1}$ and $\mathbf{y_2}$ for descriptions 1 and 2, respectively. Then, $\mathbf{y_0}$ is constructed from the interlacing of $\mathbf{y_1}$ and $\mathbf{y_2}$.  For each set, $\mathbf{\mathbf{\hat{y}_j}}, \forall j\in\{0,1,2\}$ represents their quantized versions. 
    \item \textbf{Synthesis model} ($f_{\theta}$): A Multi-Layer Perception(MLP) that creates $\mathbf{\hat{y}_j},\forall j\in\{1,2\}$ from the original image ($\theta$ represents its parameters).
    \item \textbf{Auto-regressive model} ($f_{\psi_j}$): A MLP that estimates the distribution of subsequent pixels based on previously decoded pixels for latent spaces $\mathbf{\hat{y}_j},\forall j\in\{1,2\}$ ($\psi_j$ represents its parameters).
\end{enumerate}%
In the COOL-CHIC framework, image encoding is achieved by overfitting parameters $\{\theta, \psi, \mathbf{\hat{y}}\}$ to the image characteristics, and transmission is carried out by transmitting these parameters.
INR-MDSQC takes inspiration from this, generating two descriptions $S_1:\{\theta,\psi_1,\mathbf{\hat{y}_1}\}$ and $S_2:\{\theta,\psi_2,\mathbf{\hat{y}_2}\}$. The encoding minimizes a cost function,  that takes into consideration 
the difference between the descriptions in the latent spaces. Depending on the number of received descriptions, we use either $\mathbf{\hat{y}_1}$ or $\mathbf{\hat{y}_2}$ for reconstruction. If all the descriptions are received, the interlaced $\mathbf{\hat{y}_0}$ is used for reconstruction. The images reconstructed with $f_\theta$ are denoted as $\mathbf{\hat{x}_1}$, $\mathbf{\hat{x}_2}$, and $\mathbf{\hat{x}_0}$, respectively.
The distortion metric of each reconstruction relative to the original is the Mean Squared Error (MSE) and is defined as follows:
\begin{gather}
    D_j = \dfrac{1}{\text{C}\times \text{W}\times \text{H}}\sum_{i=1}^{\text{C}\times \text{W}\times \text{H}}({\hat{x}_{i|j} - x_{i}})^{2} \label{eq:distortionDef}\\
    \nonumber \text{where \quad} \hat{x}_{i|j} \in \mathbf{\hat{x}_j} \text{\quad and \quad}\forall j\in\{0,1,2\}
\end{gather}
with C represents the number of channels, W is the image width, H is the image height, and $i$ is the position of the pixels in raster-scan order. Thus, we denote distortions in MSE of side reconstructions with $D_1,D_2$ and central reconstruction with $D_0$.

In each set $S_j$ with $j\in\{1,2\}$, the latent space is entropy-coded for efficient transmission. This requires estimating the probability distribution $p$ of each value from the unknown signal probability distribution $q$. The entropy coding algorithm can asymptotically achieve the rate of the signal's cross-entropy $H(\mathbf{\mathbf{\hat{y}_j}})$, which is given by:
\begin{equation}
H(\mathbf{\mathbf{\hat{y}_j}}) = -E_{\mathbf{\mathbf{\hat{y}_j}} \sim q}[\log_2p(\mathbf{\mathbf{\hat{y}_j}})],\ \forall j \in \{1,2\}
\label{eq:rateDef}
\end{equation}
To estimate the distribution $p$, the autoregressive model $f_{\psi_{j}}$ estimated the entropy with the input $\mathbf{\mathbf{\hat{y}_j}}$: $p_{\psi_j}=f_{\psi_j}(\mathbf{\mathbf{\hat{y}_j}})$.
Therefore, we can establish the global MDC cost function as:
\begin{align}
    \nonumber &J_{\{\lambda_{j}, \alpha\}}(\theta,\psi_j,\mathbf{\hat{y}_{j}})
    = D_0 + \alpha\sum_{j=1}^{2}{D_j} \\
    &+\sum_{j=1}^{2} \lambda_{j}(R(\mathbf{\mathbf{\hat{y}_j}})+R(\psi_j)+R(\theta))
    \quad \forall j \in \{1,2\}
    \label{eq:costfunction}
\end{align}
where $\alpha \in [0,1]$ is redundancy factor, $R(\mathbf{\hat{y}_j})$ is the rate for the latent space $\mathbf{\hat{y}_j}$ from $p_{\psi_j}$ and will be defined in section \ref{sec:autoregressive}. The rates $R(\theta)$ and $R(\psi_j)$ are estimated for the model parameters and will be described in section \ref{sec:modelquantization}. As the selected MLP is small in size, the bitrate costs for the parameters $\theta$ and $\psi_j$ are considered negligible during training and only comes into play in the post-training optimization process. Thus, the MDC cost function given by equation (\ref{eq:costfunction}) used by the training process becomes:
\begin{equation}
    J_{\{\lambda_{j}, \alpha\}}(\theta,\psi_j,\mathbf{\hat{y}_{j}}) = D_0 + \alpha\sum_{j=1}^{2}{D_j} + \sum_{j=1}^{2} \lambda_{j}R(\mathbf{\hat{y}_j})
    \quad \forall j \in \{1,2\}
    \label{eq:costfunctiontraining}
\end{equation}
and our training objective consists of minimizing the following cost function (\ref{eq:costfunctiontraining}):
\begin{mini*}|s|
{\{\theta,\psi_j,\mathbf{\hat{y}_j}\}}{J_{\{\lambda_{j},\alpha\}}(\theta,\psi_j,\mathbf{\hat{y}_j}) \quad \forall j\in\{1,2\}}
{}{}
\end{mini*}
\subsection{Multiple description synthesis model \label{sec:synthesis}}
\begin{figure}
    \centering
    \includegraphics[width = \linewidth]{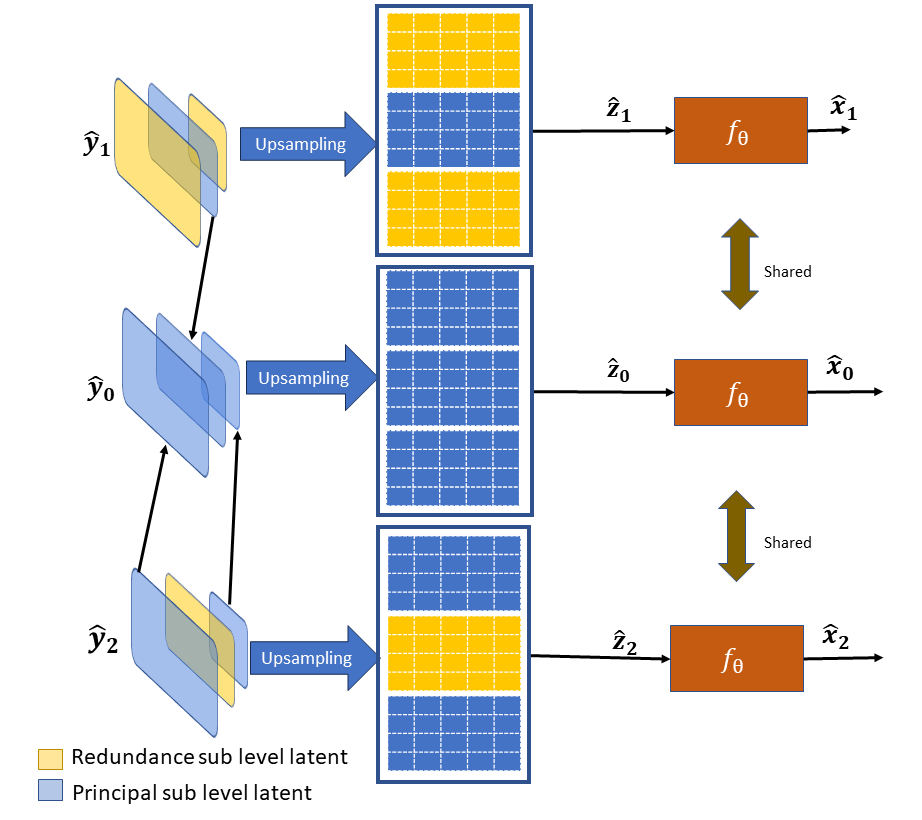}
    \caption{\footnotesize \textbf{Synthesis model}: In this example, with three decomposition levels (N=3), the central latent set $\mathbf{\hat{y}_0}$ is created by interleaving two side latent sets $\mathbf{\hat{y}_1}$ and $\mathbf{\hat{y}_2}$ as showed in the figure. Then they are upsampled to create the sets $\mathbf{\hat{z}_0},\mathbf{\hat{z}_1} \text{ \ and\ } \mathbf{\hat{z}_2} $ respectively. Finally, each up-sampled set is fed into a shared MLP for reconstruction.}
    \label{fig:Synthesismodel}
\end{figure}
First we defined the uniform scalar quantization $Q$ as:
\begin{equation}
\hat{s} = Q(s,\Delta s)
\label{eq:scalarquatizer}
\end{equation}
with s is the element to quantize and $\Delta s$ is its associated quantization step. 
We define $\mathbf{y_{k|j}}$ as the 2D latent space at level $k$ of description $j$. Each of these has a unique quantization step, and their quantized version $\mathbf{\hat{y}_{k|j}}$ is defined as:
\begin{equation*}
\mathbf{\hat{y}_{k|j}} = Q(\mathbf{y_{k|j}},\Delta \mathbf{y_{k|j}})
\end{equation*}
The quantized set of latent spaces $\mathbf{\hat{y}_j}$ for each description $j$ is defined as:
\begin{equation*}
\mathbf{\hat{y}_j} = \{\mathbf{\hat{y}_{k|j}} \in \mathbb{Z}^{H_k\times W_k}, k=0,..,N-1\}
\label{eq:hierarchic_def}
\end{equation*}
where $H_k=\dfrac{H}{2^k}$, $W_k=\dfrac{W}{2^k}$, and $N$ represents the total number of hierarchical levels of $\mathbf{\hat{y}_j}$.
When transmission is achieved without any loss of packets, we can fully receive both $\mathbf{\hat{y}_1}$ and $\mathbf{\hat{y}_2}$, and the central latent space $\mathbf{\hat{y}_0}$ is the product of interleaving between $\mathbf{\hat{y}_1}$ and $\mathbf{\hat{y}_2}$ and it is defined as:
\begin{equation*}
    \mathbf{\hat{y}_0} = \{\mathbf{\hat{y}_{2k'|1}}, \mathbf{\hat{y}_{2k'+1|2}},k'=0,1..,\nint{N / 2}\}
\end{equation*}
We design the MDC synthesis model as shown in Figure.\ref{fig:Synthesismodel},
each level of latent space will be up-sampled to the image size of $[H\times W]$ using bi-cubic interpolation. For each level in the hierarchy we have their up-sampled version:
\begin{equation*}
    \mathbf{\hat{z}_{k|j}} = \text{upsampled}(\mathbf{\hat{y}_{k|j}})
\end{equation*}
In the end, the shape of the upsampled latent space $\hat{z}_j$ is $[N\times H\times W]$. Then, the synthesis model ($f_\theta$) presents each pixel in the reconstructed image as a function of the up-sampled latent space as follows:

\begin{align*}
   \nonumber \mathbf{\hat{x}_{j}}&=f_\theta(\mathbf{\hat{z}_{j}})
   \text{ with $\mathbf{\hat{z}_{j}}=\{\mathbf{\hat{z}_{k|j}}, k=0..N-1\}$}
\end{align*}
Inspired from LMDC \cite{DBLP:journals/mta/ZhaoZBWZ22}, during training, the three up-sampled sets ${\mathbf{\hat{z}_1},\mathbf{\hat{z}_2}, \mathbf{\hat{z}_0}}$ are fed into a shared synthesis model. The network's goal is to minimize the cost function (\ref{eq:costfunctiontraining}), with the differences in distortion between the side and central reconstructions being dependent on the redundancy factor $\alpha$, which ranges from 0 to 1. The setup of the cost function (\ref{eq:costfunctiontraining}) compels the Synthesis model to partition the image information into two sets of latent descriptions, ${\mathbf{\hat{y}_1},\mathbf{\hat{y}_2}}$, under rate constraints. However, because the latent space is discrete and the quantization process isn't differentiable, uniform noise is introduced, based on \cite{DBLP:conf/pcs/BalleLS16}. This introduction of noise allows for a differentiable operation, thereby enabling gradient-based optimization. The process is detailed as:
\begin{align*}
\mathbf{\mathbf{\hat{y}_j}} &=
\begin{cases}
     \mathbf{y_j} + u, u \sim \mathcal{U}[-0.5,0.5] \quad \text{ during training}\\
     Q(\mathbf{y_j}) \quad \text{otherwise}
\end{cases}\\
\text{where \quad } &\mathcal{U} \text{: the uniform noise and } \forall j \in \{1,2\}
\end{align*}
\subsection{Auto-regressive probability model\label{sec:autoregressive}}
\begin{figure}
    \centering
    \includegraphics[width=\linewidth]{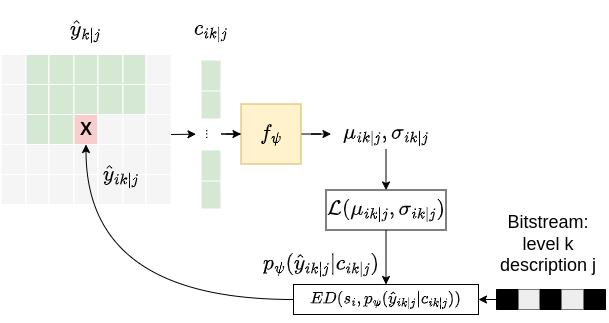}
    \caption{\footnotesize \textbf{Autoregressive model}: In this example, the model uses 12 pixels, $c_{ik|j}$, to yield ${\mu_{ik|j}}$ and ${\sigma_{ik|j}}$, modeling a Laplacian distribution. The symbol probability is calculated, and an entropy decoder estimates the latent pixel, $\hat{y}_{ik|j}$, from a bitstream.}
    

    \label{fig:autoregressive}
\end{figure}
The auto-regressive probability model so called $f_{\psi_j}$, implemented as MLP, aims to closely estimate the image's unknown latent distribution as $p_{\psi_j}$. Since the distribution of each pixel in the latent space is conditioned by their neighbor, according to \cite{DBLP:conf/nips/MinnenBT18}, the probability of the pixels is determined by a factorized model:
\begin{equation}
p_{\psi_j}(\mathbf{\hat{y}_j})=\prod_{i,k}p_{\psi_j}(\hat{y}_{ik|j}|c_{ik|j})
\label{eq:factorizedmodel}
\end{equation}
With $\hat{y}_{ik|j}$ is the latent pixel at the position $i$ of level $k$ of description $j$ and $c_{ik|j}$ are the set of decoded neighbors pixels of $\hat{y}_{ik|j}$. Therefore, $c_{ik|j} \in \mathbb{Z}^\mathcal{C}$ where $C$ is the set of causal spatially neighboring pixels.

The discrete distribution $p_{\psi_j}(\mathbf{\hat{y}_j})$ of quantized latent variables is modeled by integrating the continuous distribution of the non-quantized latent $g(y_i)$, modeled as a Laplace distribution. The MLP $f_{\psi_j}$ learns to estimate proper expectation ($\mu_{ik|j}$) and scale($\sigma_{ik|j}$) parameters for Laplacian distribution $g$ of the set of context pixels $c_{ik|j}$. Consequently, the probability of a latent pixel is modeled as:
\begin{align*}
    p_{\psi_j}(\hat{y}_{ik|j}|c_{ik|j}) = \int_{\hat{y}_{ik|j}-0.5}^{\hat{y}_{ik|j}+0.5} g(y)dy
\end{align*}
where $g \sim \mathcal{L}(\mu_{ik|j},\sigma_{ik|j})$ and $\mu_{ik|j},\sigma_{ik|j} = f_{\psi_j}(c_{ik|j})$.

As the $p_{\psi_j}$ approximates the real probability of latent space. From the article \cite{arxiv:ladune2023coolchic}, by using the factorized model equation (\ref{eq:factorizedmodel}), the rate defined in equation (\ref{eq:rateDef}) can be expressed as:

\begin{align}
    \nonumber R(\mathbf{\hat{y}_j})& = -log_2(p_{\psi_j}(\mathbf{\hat{y}_j}))= -log_2\prod_{i,k}p_{\psi_j}(\hat{y}_{ik|j}|c_{ik|j})\\
    &=-\sum_{i,k}log_2p_{\psi_j}(\hat{y}_{ik|j}|c_{ik|j})
    \label{eq:SumOfEntropy}
\end{align}
In our MDC scheme, we aim to quantize with coarser grains in redundant latent levels and finer grains in principal ones. Given the multi-resolution latent organization, the smallest resolution latent space that captures low-frequency information is more critical and must be quantized finely. Therefore, we introduce a spatial resolution coefficient $\beta_k$ to avoid excessive quantization. From equation (\ref{eq:SumOfEntropy}), our final MDC weighted rate function becomes:
\begin{equation*}
    R(\mathbf{\hat{y}_j}) =-\sum_{i,k}\beta_{k} log_2p_{\psi_j}(\hat{y}_{ik|j}|c_{ik|j}) \text{\ where\ } \beta_k = \dfrac{W_k\times H_k}{2^{2k}}
    \label{eq:estimatedRate}
\end{equation*}
\subsection{Model parameters quantization \label{sec:modelquantization}}
Compressing the INR-MDSQC model parameter consists of compressing $\{\psi_1,\psi_2,\theta\}$. During the training phase, 32-bit floating-point precision was used. However, once the training is finished, such high-precision representation is not required. {
From equation (\ref{eq:scalarquatizer}), we use three separate quantization steps, $\Delta_{\psi_1},\Delta_{\psi_2}$ and $\Delta_{\theta}$, to produce $\hat{\psi}_1,\hat{\psi}_2$ and $\hat{\theta}$, respectively. The entropy coder needs a probability model for each quantized model symbol $\hat{s}_i\subset \hat{s}$, where $\hat{s}\in\{\hat{\psi}_1,\hat{\psi}_2,\hat{\theta}\}$, in order to encode it. Empirically, the distribution of model parameters is usually best approximated by a Laplace distribution centered at 0. Therefore, we employ a Laplacian model to estimate the entropy of $\hat{s}_i$:
\begin{equation*}
p(\hat{s_i}) = \int_{\hat{s}_i-0.5}^{\hat{s}_i+0.5}g(s)ds
\end{equation*} 
where g $\sim \mathcal{L}(0,\sigma_{\hat{s}})$, $\sigma_{\hat{s}}$ is the standard deviation.
Same as function (\ref{eq:SumOfEntropy}), the estimated rate function of $\hat{s}$ can be expressed as:
\begin{align*}
    R(\hat{s}) = -\sum_{\hat{s}_i \in \hat{s}}log_2(p(\hat{s}_i)) 
\end{align*}}
{We denote $\mathbf{\hat{y}_j}$ as the fixed quantized latent space after training}. From global cost function (\ref{eq:costfunction}), our MDC post-training cost function depends only on $\hat{\theta}$ and $\hat{\psi_1},\hat{\psi_2}$:
\begin{align*}
    &J_{\lambda_{j}, \alpha}(\hat{\theta},\hat{\psi}_j) = D_0 + \alpha\sum_{j=1}^{2}{D_j}  \\&+ \sum_{j=1}^{2} \lambda_{j}(R(\mathbf{\hat{y}_j})+R(\hat{\theta})+R(\hat{\psi}_j)) \quad \forall j \in \{1,2\}
    \label{eq:posttrainingcostfunction}
\end{align*}
The minimization of the cost function above is achieved by finding the best set $\{\Delta_{\psi_1}, \Delta_{\psi_2},\Delta_\theta\}$ within a predefined range (e.g., from $10^{-1}$ to $10^{-5}$). To identify the optimal set, we independently conduct a linear search for each module. This procedure involves incrementing the quantization step sequentially and locating the corresponding step of minimum cost for each module. We apply the discovered quantization step to its respective module before moving on to the next. In our approach, we initiate this linear search with $\Delta_\theta$, followed by $\Delta_{\psi_1},\Delta_{\psi_2}$.

\subsection{Bitstream structure \label{sec:bsstructure}}
\begin{figure}[htp]
    \centering
    \includegraphics[width=\linewidth]{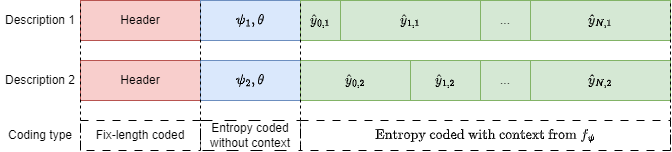}
    \caption{\footnotesize Bit-stream structure}
    \label{fig:bitstream}
\end{figure}
From the trained and quantized model, the transmitted data, formatted as depicted in Figure \ref{fig:bitstream}, starts with a header detailing the decoder's configuration parameters, including image size, layer count, model parameter quantizer steps $\Delta \psi_1,\Delta \psi_2 ,\Delta \theta$, and context pixel count. $\hat{s}_i$ where $s$ can be either $\psi_1$, $\psi_2$ or $\theta$ are entropy-coded using respective probabilities derived from $g \sim \mathcal{L}(0,\sigma_{\hat{s}})$. Finally, each latent pixel is entropy-coded using estimated probabilities from $p_{\psi_j}$ as discussed earlier.

At decoding, the header is decoded first, followed by the network parameters. Latent pixels are then decoded from the bitstream using an entropy coder initialized with source statistics estimated by the decoded auto-regressive model. Depending on the number of received descriptions, the image is reconstructed from the decoded latent space directly or via an interlacing operation between the latent space levels of the two descriptions.
\section{Experimental result \label{sec:ExpResult}}
\subsection{Implementation detail}
The framework is optimized using Synthesis MLP and Auto-regressive MLP
, both featuring two equal-sized hidden layers with 12 units each and the non linear layer used is a ReLU. Our solution is evaluated %
firstly with two images extracted from the SET4 \footnote{\url{https://github.com/mdcnn/MDCNN_test40/tree/master/SET4}} dataset, Lena (4.png) and the boat (1.png), and secondly, with all the images from the DIV2k \footnote{\url{https://github.com/mdcnn/MDCNN_test40/tree/master/DIV2K-10}} dataset. We set the number of hierarchical levels to 6 for Lena and boat images, and to 8 for the DIV2K dataset. The Adam optimizer is used for optimization, with an initial learning rate of $l_r=0.1$ and 10000 iterations. For the high-resolution DIV2K dataset, we only tested with $\alpha=0.1$, while for the other datasets, we tested with $\alpha=0.1$ and $\alpha=1.0$. The distortion is estimated through Peak Signal-to-Noise Ratio (PSNR) and Multi-Scale Structural Similarity Index (MS-SSIM) \cite{DBLP:journals/tip/WangBSS04}, while bit-per-pixel (bpp) is used to determine the compression rate. The entropy codec used is Range Coder from \cite{bamler2022constriction}.

\begin{figure*}
    \centering
    \begin{minipage}{0.25\textwidth}
        \centering
        \includegraphics[width=\linewidth]{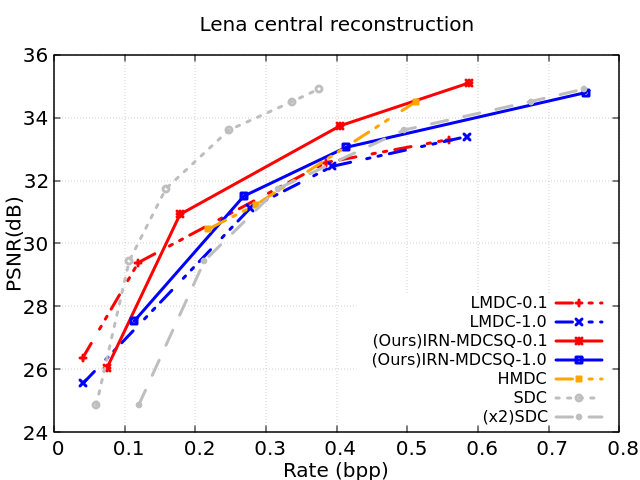}
    \end{minipage}%
    \begin{minipage}{0.25\textwidth}
        \centering
        \includegraphics[width=\linewidth]{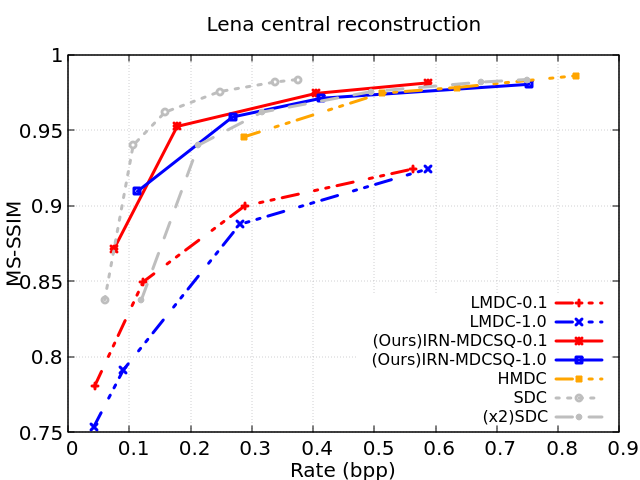}
    \end{minipage}%
    \begin{minipage}{0.25\textwidth}
        \centering
        \includegraphics[width=\linewidth]{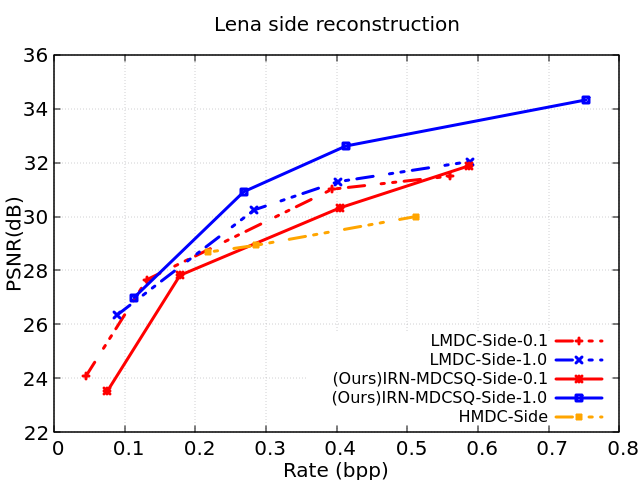}
    \end{minipage}%
    \begin{minipage}{0.25\textwidth}
        \centering
        \includegraphics[width=\linewidth]{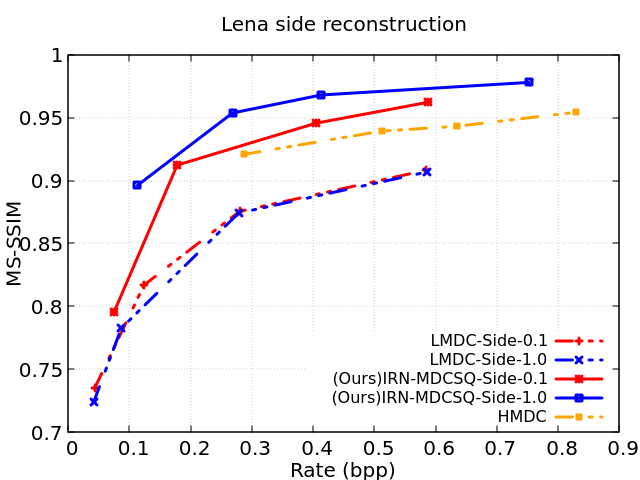}
    \end{minipage}
    \begin{minipage}{0.25\textwidth}
        \centering
        \includegraphics[width=\linewidth]{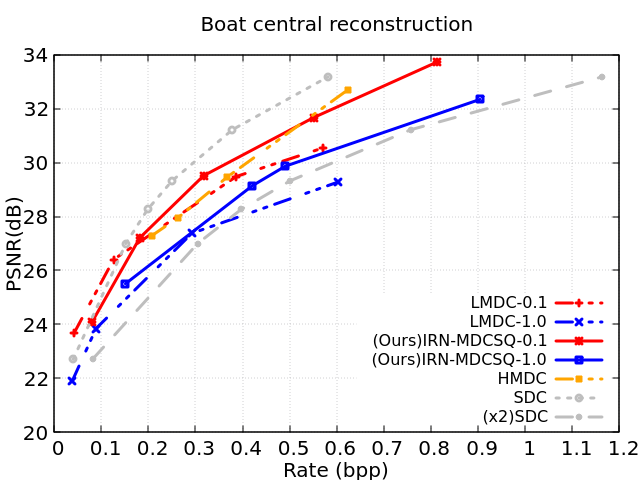}
    \end{minipage}%
    \begin{minipage}{0.25\textwidth}
        \centering
        \includegraphics[width=\linewidth]{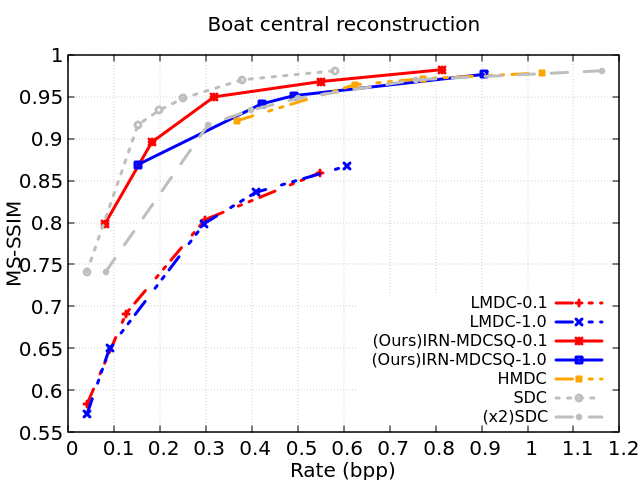}
    \end{minipage}%
    \begin{minipage}{0.25\textwidth}
        \centering
        \includegraphics[width=\linewidth]{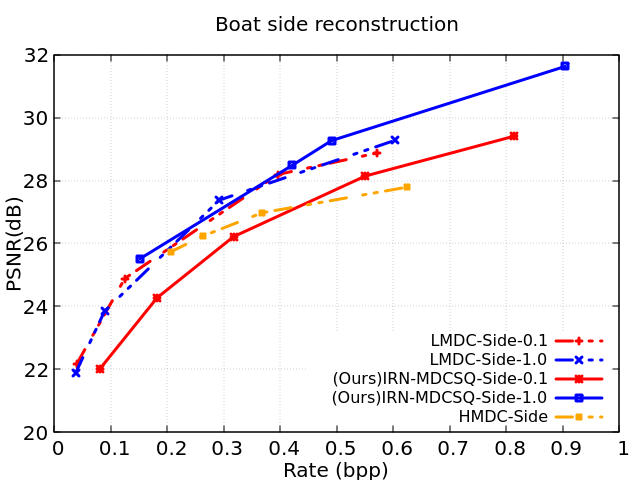}
    \end{minipage}%
    \begin{minipage}{0.25\textwidth}
        \centering
        \includegraphics[width=\linewidth]{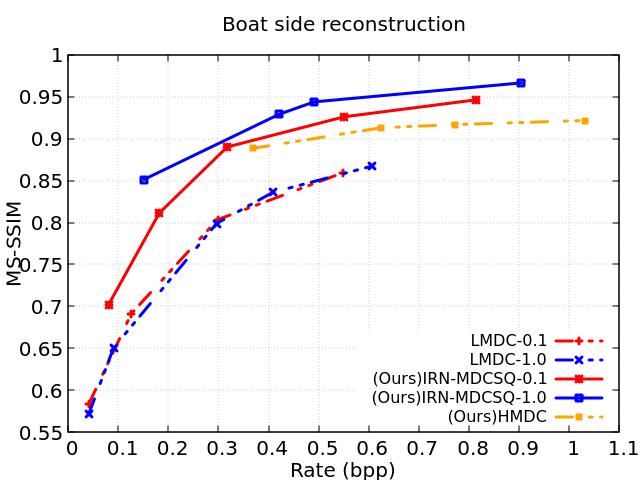}
    \end{minipage}
    \begin{minipage}{0.25\textwidth}
        \centering
        \includegraphics[width=\linewidth]{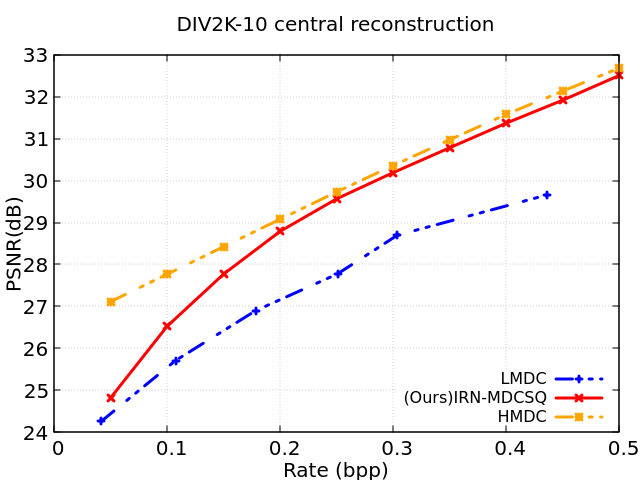}
    \end{minipage}%
    \begin{minipage}{0.25\textwidth}
        \centering
        \includegraphics[width=\linewidth]{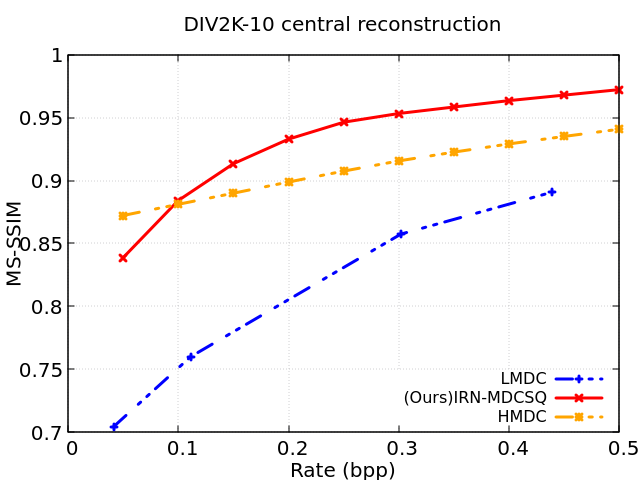}
    \end{minipage}%
    \begin{minipage}{0.25\textwidth}
        \centering
        \includegraphics[width=\linewidth]{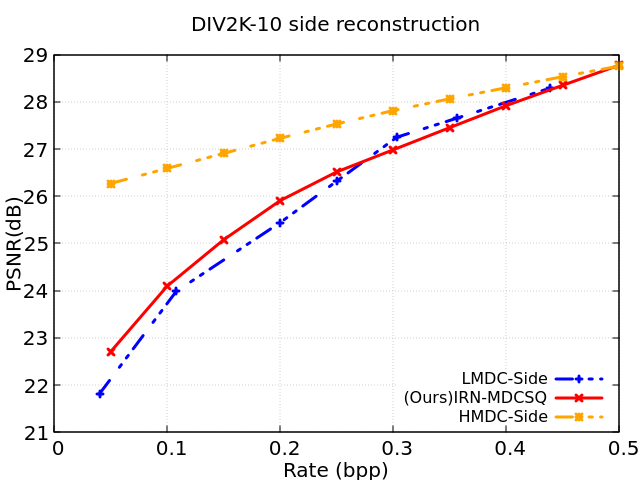}
    \end{minipage}%
    \begin{minipage}{0.25\textwidth}
        \centering
        \includegraphics[width=\linewidth]{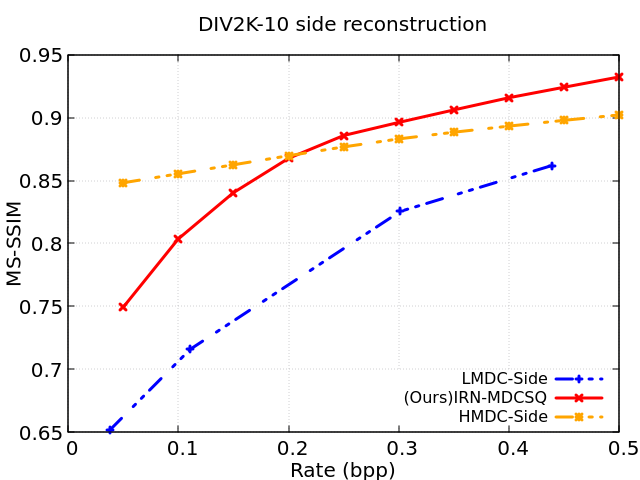}
    \end{minipage}
    \caption{\footnotesize The rate distortion performance was evaluated on Lena and Boat images (512x512 pixels) for two values of redundancy ($\alpha=0.1$ and $\alpha=1$) and the The DIV2K-10 dataset (1920x1080 px) for one redundancy factor ($\alpha=0.1$) }
    \label{fig:result_RD}
\end{figure*}

The image's central reconstruction is the result of the interlacing process between two descriptions of different hierarchical levels, hence the quantity of information captured in the latent spaces $\mathbf{\hat{y}_1}$ and $\mathbf{\hat{y}_2}$ is unequal. Indeed, because the way the two descriptions are defined in the latent space, the proposed MDC is unbalanced. Therefore, for the same value of $\lambda_j$ between two descriptions, the description which contain the lowest latent resolution as the principal latent will exhibit higher quality than the other. This property is beneficial because each description is transmitted over distinct, independent channels, each one possessing unique characteristics. The description of higher quality is dispatched via a less noisy channel, while the description of lower quality is conveyed through a channel with more noise.
Due to the unequal rate, there is a consequent unequal distortion between the two descriptions. The side distortion curve showed in the figure \ref{fig:result_RD} represents the PSNR of the average of the MSE across all side descriptions.

To ensure the validity of our solution, the performance of our INR-MDSQC at central reconstruction should not surpass the upper limit of the SDC set by the original coder COOL-CHIC, nor should it fall below the SDC at double rate. As showcased in the Lena and boat images (see Figure \ref{fig:result_RD}), with $\alpha=0.1$, the solution approaches the upper bound limit of the single SDC. When full redundancy is applied, $\alpha=1.0$, the performance of our MDC coincides with the lower limit SDC at double rate.%
\subsection{Rate distortion study}
We benchmarked our solution against LMDC \cite{DBLP:journals/mta/ZhaoZBWZ22}. Additionally, we compared our approach with classic MDC methods such as HEVC-MDC (HMDC)\cite{le2023multiple}. The results for LMDC can be found in \cite{DBLP:journals/mta/ZhaoZBWZ22}, and the results for HMDC were obtained from a re-implementation of the method.

\begin{figure*}
    \centering{(a) INR-MDSQC}\\
    \begin{minipage}{0.25\textwidth}
        \centering
        \includegraphics[width=0.95\linewidth]{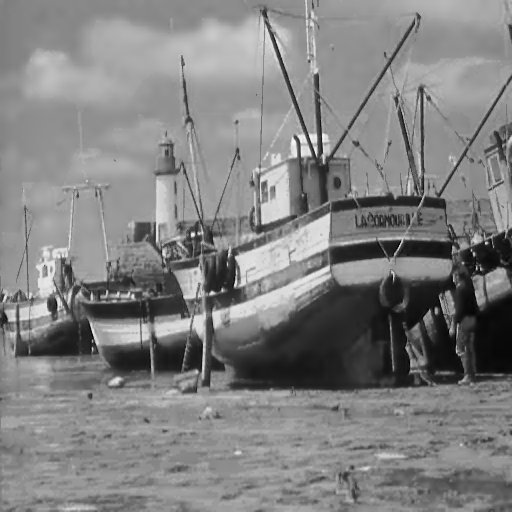}
        \centerline{Central: 31.79dB-0.54bpp}
    \end{minipage}%
    \begin{minipage}{0.25\textwidth}
        \centering
        \includegraphics[width=0.95\linewidth]{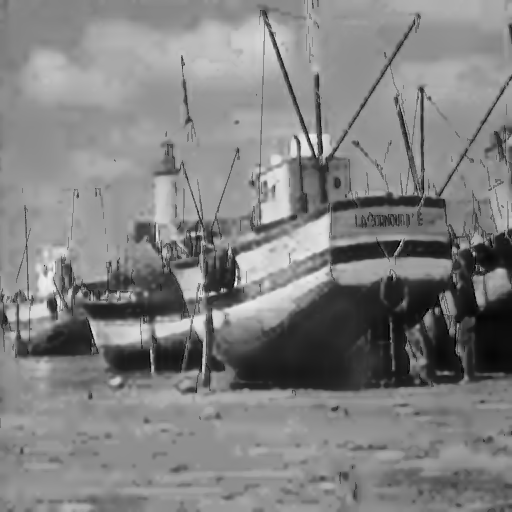}
        \centerline{Side1: 27.15dB-0.21bpp}
    \end{minipage}%
    \begin{minipage}{0.25\textwidth}
        \centering
        \includegraphics[width=0.95\linewidth]{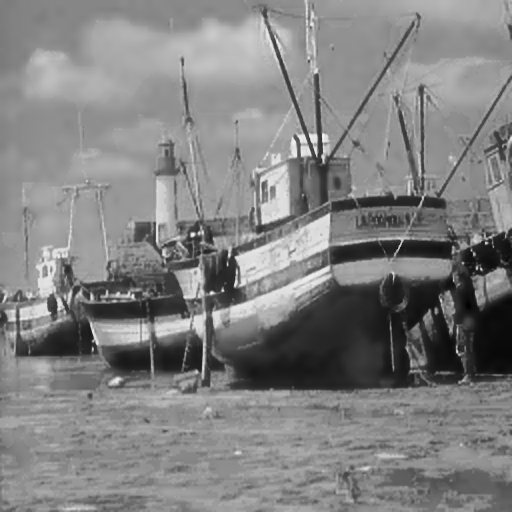}
        \centerline{Side2: 28.95dB-0.33bpp}
    \end{minipage}
    \centerline{(b)HMDC}
    \begin{minipage}{0.25\textwidth}
        \centering
        \includegraphics[width=0.95\linewidth]{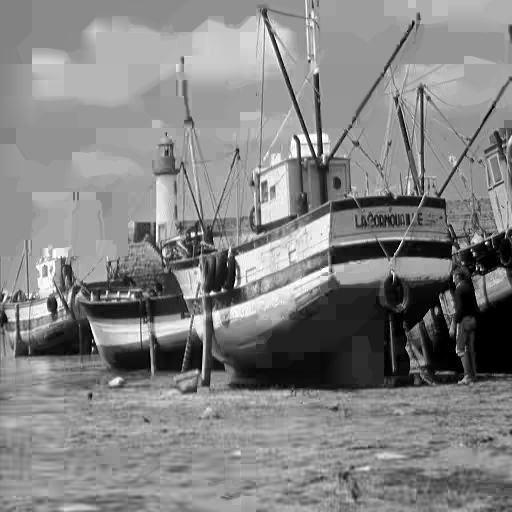}
        \centerline{Central:31.25dB-0.58bpp}
    \end{minipage}%
    \begin{minipage}{0.25\textwidth}
        \centering
        \includegraphics[width=0.95\linewidth]{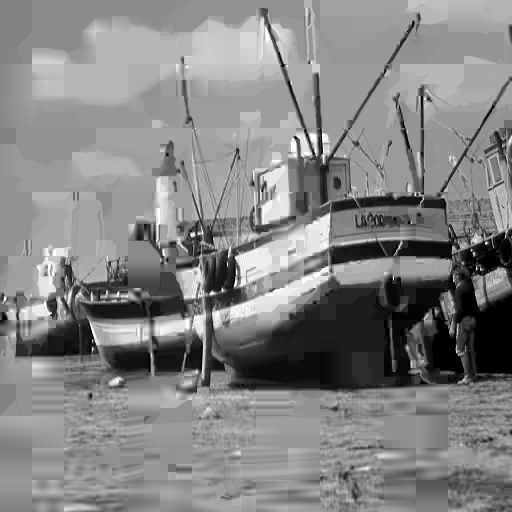}
        \centerline{Side1:27.63dB-0.27bpp}
    \end{minipage}%
    \begin{minipage}{0.25\textwidth}
        \centering        
        \includegraphics[width=0.95\linewidth]{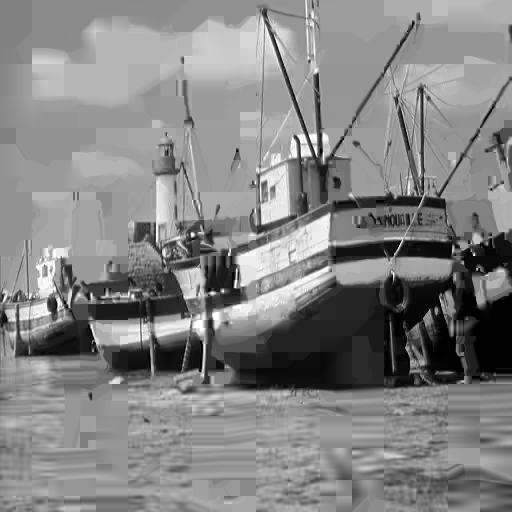}
        \centerline{Side2:27.96dB-0.31bpp}
    \end{minipage}
    \caption{ \footnotesize The visual results  of the image "Boat", shows that the blocking artifact caused by high quantization is more noticeable on the HMDC than INR-MDSQC at a "nearly" identical bitrate.}
    \label{fig:VisualizationofBoat}
\end{figure*}

With the small images of lena and of the boat from the SET4 
, with redundancy $\alpha=0.1$, our solution shows an improvement in PSNR at high bit-rates compared to LMDC and overperform the HMDC method. This can be attributed to the fact that, at low bit-rates, the cost of coding $\theta,\psi_j$ becomes significant. However, at high bit-rates, our solution adapts more effectively to the image characteristics, thus enhancing the quality of the reconstruction. In terms of the MS-SSIM metric, our method also surpasses the LMDC and HMDC methods. Figure \ref{fig:VisualizationofBoat} presents a comparison of the visual outcomes from HMDC and our solution. Our method tends to produce less blocking artifacts. When full redundancy ($\alpha = 1.0$) is applied, our method achieves higher performance in side reconstruction, with improved central reconstruction performance compared to LMDC

In the high-resolution dataset DIV4K-10, traditional MDC strategies such as HMDC still outperform INR-MDSQC methods in terms of PSNR for side reconstruction. However, INR-MDSQC achieves nearly identical PSNR performance for central reconstruction. Moreover, in terms of MS-SSIM, our solution outperforms the HMDC method. Importantly, our method maintains superior central reconstruction performance compared to LMDC while achieving similar distortion levels for side reconstruction.
\section{Conclusion \label{sec:conclusion}}
We introduce INR-MDSQC, an Implicit Neural Representation Multiple Description Scalar Quantizer Codec, which is built based on the COOL-CHIC framework. By overfitting the neural network for each image, INR-MDSQC can capture more details, thereby enhancing performance compared to the traditional Autoencoder MDC approach. Furthermore, this framework allows for more flexible redundancy tuning. When compared to conventional MDC frameworks, our solution delivers superior reconstruction quality in term of MS-SSIM at almost the same central PSNR. From our perspective, a study aimed at reducing complexity is necessary to enhance the method's efficiency. Moreover, an evaluation of the system's performance under noisy channel conditions will be required in our future works.
\subsection*{Acknowledgement}
We would like to extend our gratitude to Jérémy Mateos for his innovative ideas, suggestions that improved the performance of our model.

\bibliographystyle{IEEEbib}
{\footnotesize
\bibliography{refs}
}
\end{document}